\documentclass{article}
\usepackage{spconf,amsmath,graphicx}
\usepackage{mathrsfs}
\usepackage{times,amsmath,epsfig}
\usepackage{bm}
\usepackage{amsfonts}
\usepackage[numbers,sort&compress]{natbib}
\emergencystretch=\maxdimen
\begin{document}\sloppy
\def\x{{\mathbf x}}
\def\L{{\cal L}}

\title{Deep Multiple Description Coding by Learning \\Scalar Quantization}
%
\name{Lijun Zhao{\small $~^{\star}$},Huihui Bai{\small $~^{\star}$}, Anhong Wang{\small $~^{\#}$}, Yao Zhao{\small $~^{\star}$}
\thanks{Corresponding author: Huihui Bai (email: hhbai@bjtu.edu.cn). This work was supported in part by National Natural Science Foundation of China (No. 61672087, 61672373) and the Fundamental Research Funds for the Central Universities (K17JB00260, K17JB00350).}}

\address{{\small $~^{\star}$}Institute Information Science, Beijing Jiaotong University, P. R. China\\
{\small $~^{\#}$}Institute of Digital Media and Communication, Taiyuan University of Science and Technology}

%
%
%

%
\maketitle
\begin{abstract}
In this paper, we propose a deep multiple description coding framework, whose quantizers are adaptively learned via the minimization of multiple description compressive loss. Firstly, our framework is built upon auto-encoder networks, which have multiple description multi-scale dilated encoder network and multiple description decoder networks. Secondly, two entropy estimation networks are learned to estimate the informative amounts of the quantized tensors, which can further supervise the learning of multiple description encoder network to represent the input image delicately. Thirdly, a pair of scalar quantizers accompanied by two importance-indicator maps is automatically learned in an end-to-end self-supervised way. Finally, multiple description structural dissimilarity distance loss is imposed on multiple description decoded images in pixel domain for diversified multiple description generations rather than on feature tensors in feature domain, in addition to multiple description reconstruction loss. Through testing on two commonly used datasets, it is verified that our method is beyond several state-of-the-art multiple description coding approaches in terms of coding efficiency.
\end{abstract}
\section{Introduction}
Traditional multiple description coding (MDC) approaches have been widely studied in the last decades, for which the derivation of multiple description theoretical rate-distortion regions is a fundamental and significant topic. Meanwhile, the achievable rate-distortion regions in practice gradually approach the boundaries of theoretical rate-distortion regions of multiple description coding. However, a large number of traditional multiple description coding approaches face many thorny problems. For example, multiple description quantizers often need to assign the optimal index for multiple description generations, especially quantizing for more than two descriptions, which is an extremely complicated problem. For quantization-based MDC methods, there are mainly three classes: scalar quantizers, trellis coded quantizers, and lattice vector quantizers. In the early time, multiple description scalar quantizers constrained by the symmetric entropy are formed as an optimization problem \cite{qq27}. When generalizing this method from two to L descriptions during the optimization of the encoder, linear joint decoders are developed to resolve the problem of dynastical computation increase \cite{qq17}. In \cite{qq28}, the distortion-rate performance is derived for certain randomly-generated quantizers. By generalizing randomly-generated quantizers, the theoretical performances of MDC with randomly offset quantizers are given in the closed-form expressions \cite{qq13}, while a lower bound is achieved for multiple description uniformly offset. To increase the robustness to bit errors, linear permutation pairs are developed for index assignment for two description scalar quantizers \cite{qq10}. Unlike scalar quantizers, a trellis is formed by the tensor product of trellises for multiple description coding \cite{qq26}. Built upon on this work, two-stage MD image coding method uses course quantization to get the shared information so as to explicitly adjust multiple description redundancy, after which the trellis coded quantization is used for second-stage coding \cite{qq6}.

Because scalar quantizers and trellis coded quantizers always require complicated index assignment, lattice vector quantizers have been explored extensively in the last decades due to its many advantages such as inherent symmetry structure, no need to design codebook and avoiding complex nearest neighbor search, etc.  In \cite{qq14}, the lattice vector quantizer design problem is cast into a labeling problem, and a systematic construction method is detailed for general lattices. To further improve the performance of multiple description lattice vector quantization at the cost of little complexity increase, the fine lattice codebook is replaced by a non-lattice codebook \cite{qq15}. When one description is correctly received, but the other has some bit-error \cite{qq18}, a structured bit-error resilient mapping is leveraged to make multiple description lattice vector quantizers resilient to the bit-error by exploiting lattice's intrinsic structural characteristic. To be capable of balancing different received descriptions reconstruction quality, a heuristic index assignment algorithm is given to control the distortions and a different reconstruction method is used for $L \geq 3$ descriptions coding \cite{qq7}.

For correlating transform-based MDC framework, a good performance can be achieved for multiple description coding with a small amount of redundancy between different descriptions, but this framework's coding efficiency always will decrease if more redundancy is introduced. Unlike multiple description quantizer and correlating transform-based MDC framework, sampling-based multiple description coding is more flexible and can be compatible with standard coders. But most of the current sampling-based MDC methods always manually design a specific sampling method based on their empirical knowledge or extend the existing sampling methods for multiple description generations. The coding efficiency of these MDC methods is limited by non-adaptive sampling. Consequently, the sampling-based MDC methods require to be further improved. Recently, a convolutional neural network (CNN) based MDC method \cite{qq11} tries to adaptively sample the input image to create multiple descriptions, but its coding efficiency is still not high enough. In summary, we should further study the topic of extremely compressing images at low bit-rate with multiple description coding for error resilience against bit-errors and package loss.

In this paper, a deep multiple description coding framework is proposed to compress images for robust transmission. There are several main parts: multiple description multi-scale dilated encoder network, multiple description scalar quantization, context-based entropy estimation neural networks, and multiple description decoder networks. Our contributions are listed below: (1) a general deep multiple description coding framework is built upon convolutional neural networks; (2) a pair of scalar quantization operators is automatically learned in an end-to-end self-supervised way to generate multiple diverse descriptions; (3) each scalar quantizer is accompanied by an importance-indicator map to better quantize the feature tensors; (4) we propose to use multiple description structural dissimilarity distance loss for multiple description decoded images, which implicitly regularizes diversified multiple description generations and scalar quantization's learning, but there is no distance loss imposed on feature tensors generated by multiple description encoder network.

\section{Deep Multiple Description Coding Framework}

\begin{figure*}[!hbt]
\centering
\includegraphics[width=6in]{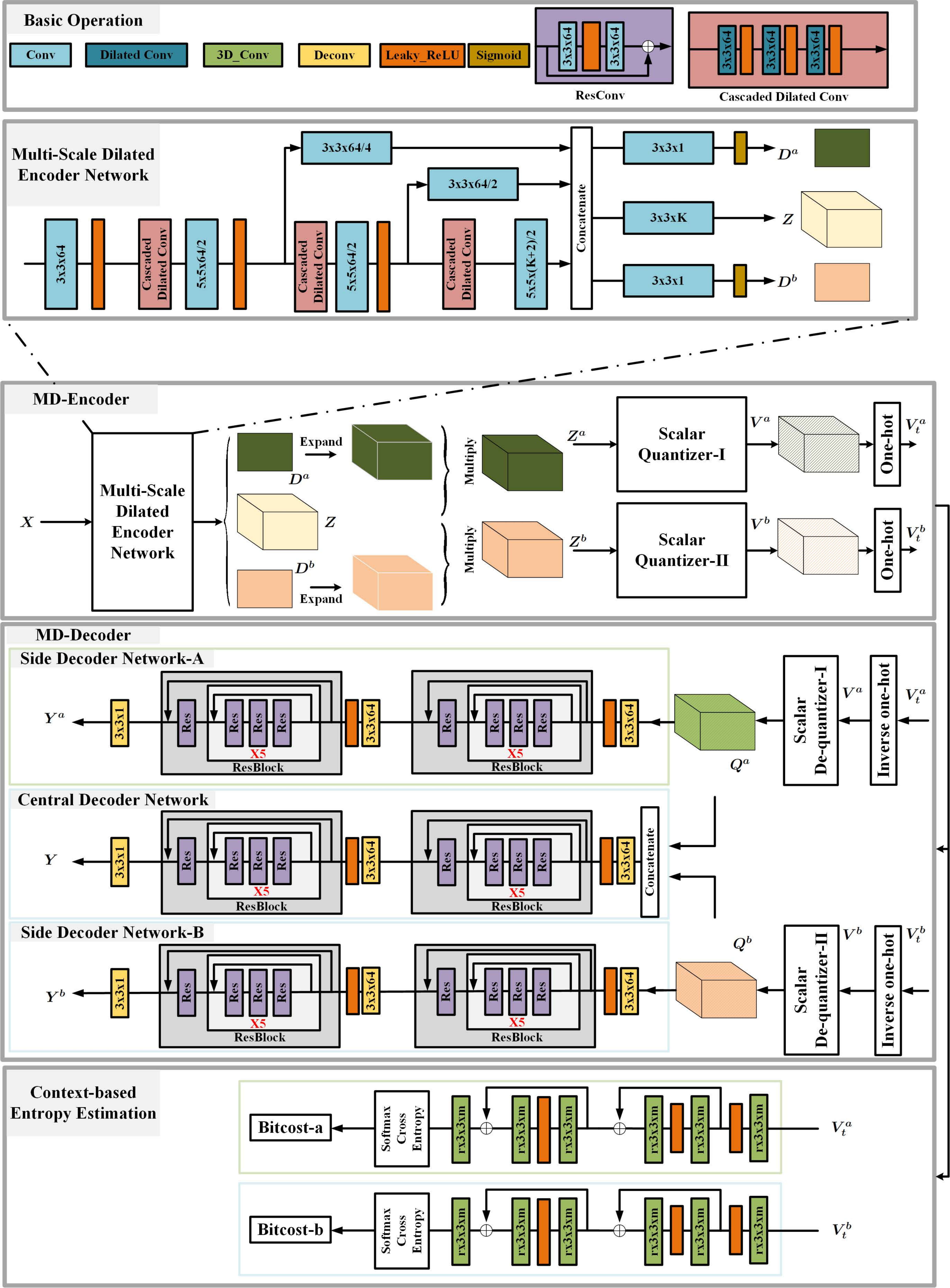}
\caption{The diagram for deep multiple description coding framework. (Note that 3x3x64 means that spatial kernel size is 3x3, the number of output feature maps is 64, and the stride is 1 in this convolutional layer in default, while 3x3x64/2 shares similar denotation except for with a stride of 2. Meanwhile, other convolutional layers can be denoted similarly. One-hot refers to the one hot operation.)}
\label{Fig1}
\end{figure*}

In this paper, we introduce a deep multiple description coding framework, which is entirely built upon deep convolutional neural networks, as displayed in Fig. \ref{Fig1}. Our framework is mainly composed of MD encoder, MD decoder, MD scalar quantizers, and two context-based entropy estimation networks. Given an input image $\bm{X}$, multi-scale dilated encoder network in the MD encoder decomposes this image into a feature tensor $\bm{Z}$, and two importance-indicator maps $\bm{D^a}$ and $\bm{D^b}$. Before scalar quantization, the expansion of each importance-indicator map is multiplied by the feature tensor $\bm{Z}$ so as to obtain two new feature tensors $\bm{Z^a}$ and $\bm{Z^b}$. Here, the expansion operation is done by following the work of M.~Fabian and co-workers \cite{qq22}. Then, these two feature tensors $\bm{Z^a}$ and $\bm{Z^b}$ are respectively quantized by scalar quantizer-I and scalar quantizer-II. Due to the non-differentiability of hard quantization, we follow the work of \cite{qq41} to utilize the soft quantization to make our framework capable of being trained in an end-to-end way. The scalar quantizer-I is represented as:
\begin{align}
&v^a_i=\mathop{\arg\max}_{c_j\in \bm{C^a}} \zeta(-\sigma ||z^a(i)-c_j||^2), j \in [1,2,...,n],\notag\\
&\zeta(-\sigma ||z^a(i)-c_j||^2)=\frac{e^{-\sigma ||z^a(i)-c_j||^2}}{\sum_{k=1}^{K} e^{-\sigma ||z^a(i)-c_k||^2}},
\end{align}
where $||\cdot||$ denotes $L2$-norm, $\bm{C^a}=[c_1,...,c_n]$ is the center variable vector, and $z^a(i)$ represents the $i$-th element of tensors $\bm{Z^a}$.

According to Eq.(1), the scalar quantizer-II with the center variables $\bm{C^b}$ can be defined in this way. The vectors of quantized tensors $\bm{V^a}$ and $\bm{V^b}$ can be represented as $[v^a_1,..., v^a_n]$ and $[v^b_1,..., v^b_n]$. This pair of scalar quantizers accompanied by two importance-indicator maps can be learned in an end-to-end self-supervised way to generate multiple diverse descriptions. Both of tensors $\bm{V^a} \in \mathcal{N}^{M \times N \times K}$ and $\bm{V^b}\in \mathcal{N}^{M \times N \times K }$ can be converted as the one-hot tensors $\bm{V_t^a}\in \mathcal{N}^{M \times N \times S \times T} $ and $\bm{V_t^b}\in \mathcal{N}^{ M \times N \times S \times T }$ by adding a new dimension, which are the two generated descriptions. The generated descriptions are losslessly encoded and transmitted over channels. During forward propagation, we use hard quantization according to Eq.(1), but we use the derivation of soft quantization function \cite{qq41} to back-propagate gradients from the MD decoder networks to MD encoder network.

At the receiver, the decoded one-hot tensors $\bm{V_t^a}$ and $\bm{V_t^b}$ can be reversibly converted into the tensors $\bm{V^a}\in \mathcal{N}^{M\times N\times K}$ and $\bm{V^b}\in \mathcal{N}^{M\times N\times K}$. After that, these tensors are dealt by corresponding scalar de-quantizers. The scalar de-quantizer-I can be written as : $\bm{Q^a}(i)= \hbar (\bm{V^a}(i))$, which returns the $\bm{V^a}(i)$-th of center variable $\bm{C^a}$. Similarly, scalar de-quantizer-II can be written as $\bm{Q^b}(i)= \hbar (\bm{V^b}(i))$. The side decoder network-A or side decoder network-B is used to decode the quantized tensor $\bm{Q^a}$ or $\bm{Q^b}$ as the lossy image $\bm{Y^a}$ or $\bm{Y^b}$, when only one description is received in the decoder, which is transmitted over an unpredictable channel. If both of these descriptions are received, we leverage central decoder network to decode images with the quantized tensors $\bm{Q^a}$ and $\bm{Q^b}$, as shown in Fig. \ref{Fig1}.

\subsection{The objective function of multiple description coding}
Just as traditional single description image compression, the objective function of MD coding requires to balance two fundamental parts: coding bitrate and MD image decoding distortion. Traditional single description image compression always employs mean square error (MSE) to measure image compression distortion. However, the human visually perceptual quality of the compressed image with high MSE may be higher than the one with low MSE \cite{qq20}. For objective-subjective quality mismatching, there are a great many reasons, such as one-pixel position shifting of the whole-image, new-pixels covered over details-missing textural regions generated by generative adversarial networks for a sense of reality. Compared to MSE loss, mean absolute error (MAE) loss can better regularize image compression to make the compressed images to move towards the ground-truth images during training. Thus, our framework use MAE loss for both side decoded images and central decoded image as the first part of our multiple description reconstruction loss, which can be written as follows:
\begin{align}
&D_{L1}(\bm{X},\bm{Y^a},\bm{Y^b},\bm{Y})=\frac{1}{64\cdot M \cdot N}\sum_{i}(||\bm{X}_i-\bm{Y^a}_i||_1)+ \notag\\
&\frac{1}{64\cdot M \cdot N}\sum_{i}(||\bm{X}_i-\bm{Y^b}_i||_1)+\notag\\
&\frac{1}{64\cdot M \cdot N}\sum_{i}(||\bm{X}_i-\bm{Y}_i||_1).
\end{align}
Meanwhile, to better measure image distortion for structural preservation, we introduce multi-resolution structural similarity index (MR-SSIM) $ f_{MR}(\cdot, \cdot)$ as an image quality evaluation factor according to \cite{qq21}. Image's distortions at different scales are of very different importance regarding perceived quality. The MR-SSIM's weights $[0.750, 0.188, 0.047, 0.012, 0.003]$ for different scales are linearly proportional to image size for each-scale image, that is to say, large-scale image's weight is bigger than small-scale one. Different to our MR-SSIM, MS-SSIM in literature \cite{qq21} uses the weight of $[0.0448, 0.2856, 0.3001, 0.2363, 0.1333]$. This weight means that other-scale images are more significant than the largest and smallest scale images. The total structural dissimilarity loss as the second part of our multiple description reconstruction loss can be written as:
\begin{align}
&D_{\rm{{MR}}} (\bm{X}, \bm{Y^a}, \bm{Y^b}, \bm{Y}) =- f_{MR}(\bm{X}, \bm{Y^a})\notag\\
& - f_{MR}(\bm{X}, \bm{Y^b})- f_{MR}(\bm{X}, \bm{Y}).
\end{align}

Unlike single description image compression with only one bit-stream produced by the coder, multiple description coding should generate multiple diverse descriptions, between which some redundancy is shared, but each description has its unique information. The redundancy of these description makes the receiver capable of decoding an acceptable quality image, even though one description is missing when multiple description bit-streams are transmitted over the unstable channel. However, when different descriptions have too much-shared information, the central image quality does not have great improvements, even though all the descriptions are got at the client. In \cite{qq11}, multiple description distance loss is directly used to supervise multiple description generations in the image space. Although the feature tensor of each description can be used as the opposite label to regularize each other, the problem of multiple description leaning on feature space is often a very tricky problem, because the same multiple description reconstruction may come from the composition of different features. To let our framework to automatically generate multiple diverse descriptions, we propose to use multiple description structural dissimilarity loss in our multiple description distance loss. Multiple description distance loss explicitly regularizes multiple description decoded images to be different, and implicitly regularize scalar quantization's learning and diversified multiple description generations, which is written as:
\begin{equation}
D_{distance}=f_{MR}(\bm{Y^a}, \bm{Y^b}).
\end{equation}
To precisely predict each description's coding costs, we use two entropy estimation networks without network's parameter sharing. Follow the work of \cite{qq22}, we use context-based entropy estimation neural networks to efficiently estimate each description's coding costs. The estimated coding costs for each description are respectively denoted as $R(\bm{V^a})$ and $R(\bm{V^b})$. Finally, our framework's multiple description compressive loss can be written as:
\begin{align}
&f(\bm{X}, \bm{Y^a}, \bm{Y^b}, \bm{Y})=[D_{L1}(\bm{X},\bm{Y^a},\bm{Y^b},\bm{Y})+ \notag\\
&D_{\rm{{MR}}}(\bm{X}, \bm{Y^a},\bm{Y^b}, \bm{Y})]+\alpha D_{distance} + \beta D_{reg}\notag\\
&+ \gamma [R(\bm{V^a})+ R(\bm{V^b})],
\end{align}
in which $D_{reg}$ is the regularization term for the parameters of convolutional neural networks. And $\alpha$, $\beta$, and $\gamma$ are three hyper-parameters.

\subsection{Network}
To fully explore image context information, we propose a multiple description multi-scale dilated encoder network to get a feature tensor $\bm{Z}$ and two importance-indicator maps $\bm{D^a}$ and $\bm{D^b}$ for multiple description generation, as shown in Fig.1. As discussed in \cite{qq23}, dilated convolution can extremely enlarge image receptive field, but it may introduce the gridding effect. In \cite{qq24}, a three-layers cascaded dilated convolution is defined as hybrid dilated convolution to resolve this problem for semantic segmentation. Therefore, we use three cascaded dilated convolutions to extract multi-scale features, since it is vital to leverage image context information for diversified multiple description generations. As shown in Fig. \ref{Fig1}, one convolutional layer is used before three cascaded dilated convolutions, while each cascaded dilated convolution is followed by $5\times5$ down-sampling convolutional layer with a stride of 2. To leverage image multi-scale context information, the first cascaded dilated convolution is followed by a down-sampling convolutional layer with a stride of 4. Meanwhile, a convolutional layer with a stride of 2 is used to down-sample the second cascaded dilated convolution's output features in the spatial domain. After that, various features from different scales are concatenated, and then aggregated together by three $3\times3$ convolution layers to get a feature tensor $\bm{Z}$ and two importance-indicator maps $\bm{D^a}$ and $\bm{D^b}$ respectively, as depicted in Fig.1.

After receiving de-quantized tensors $\bm{Q^a}$ and $\bm{Q^b}$, we use multiple description decoder networks to decode these tensors. From Fig.1, it can be found that side decoder networks and central decoder network share a similar network structure except for different inputs. When all the multiple descriptions are received, both of de-quantized tensors $\bm{Q^a}$ and $\bm{Q^b}$ are concatenated and fed into the central decoder network for image decoding. If one description is missing, but the other description is got, side decoder network-A or side decoder network-B is chosen to decode the de-quantized tensor $\bm{Q^a}$ or $\bm{Q^b}$. These networks are composed of three deconvolution layers and two ResBlocks. Each one of first two deconvolution layers is followed by one ResBlock. We use 16 ResConv with skip-connection in each ResBlock, as depicted in Fig. \ref{Fig1}. Besides, two entropy estimation networks have the same network structure in \cite{qq22}, which has six 3D-convolution layers, as shown in Fig. \ref{Fig1}. The middle four convolution layers in the structure are cascaded with two skip-connection, which build up two 3D-Resconv.
\begin{figure*}[!htb]
\centering
\includegraphics[width=6in]{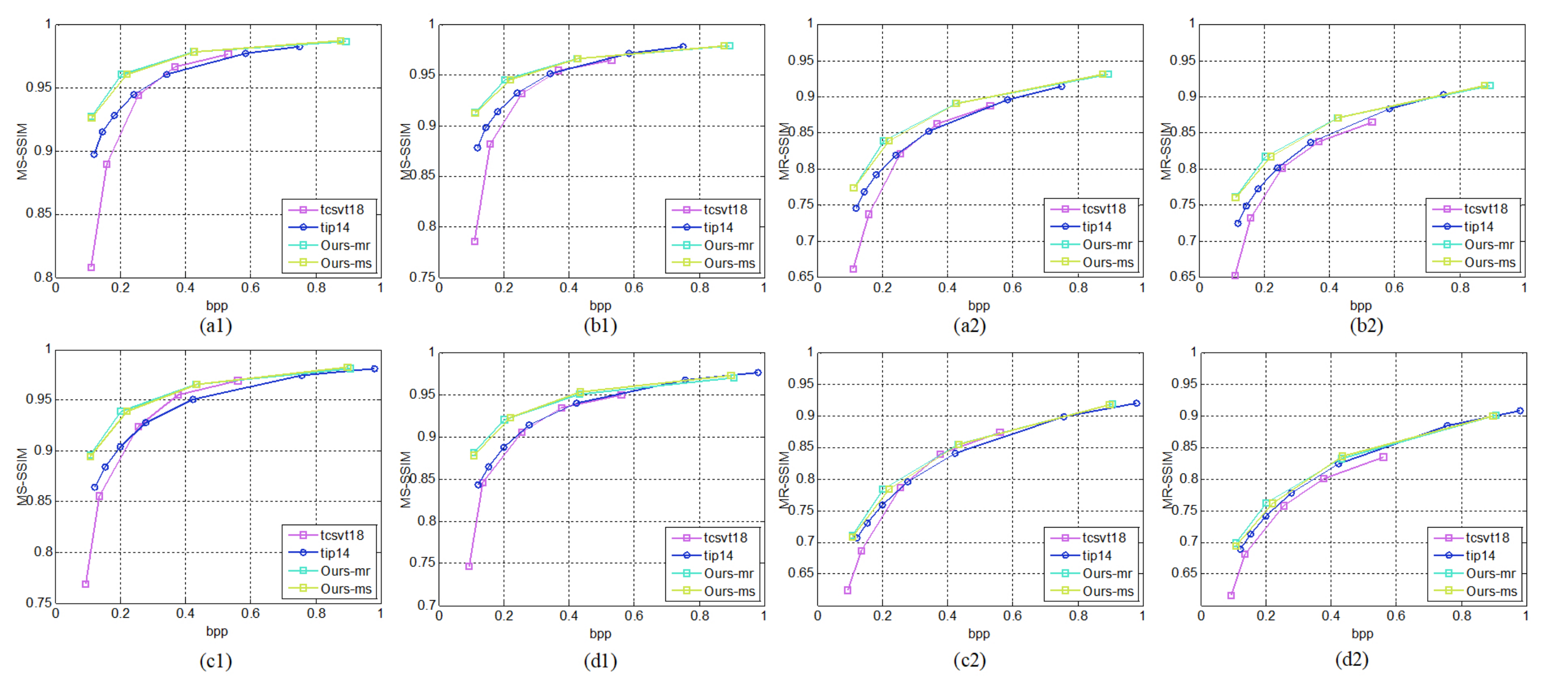}
\caption{The average objective quality comparisons of different multiple description coding methods. (a1-b1) and (a2-b2) are respectively the results of MS-SSIM and MR-SSIM for decoded central and side images testing on Set4, (c1-d1) and (c2-d2) are respectively the results of MS-SSIM and MR-SSIM for decoded central and side images testing on Kodak PhotoCD dataset}
\label{Fig2}
\end{figure*}

\section{Experimental results}

\subsection{Training details}
We train our framework on the training dataset of ImageNet from ILSVRC2012. During training, each image patch with the size of $160 \times 160$ is got by randomly cropping the training images, but we will first resize the input image to be at least 160 for each dimension before cropping, if the training image size is smaller than $160 \times 160$. Moreover, ImageNet's validation dataset is used as our validation datasets. Two datasets are chosen as our testing dataset for the comparison of different MDC methods. The first dataset is Set4 used in \cite{qq11}, while the second dataset is the Kodak PhotoCD dataset with a size of $768 \times 512$, which is a commonly used testing dataset for image compression. During training, we choose Adam optimization to minimize the objective loss of our MDC framework with the initial learning rate to be 4e-3 for auto-encoder networks. The training batch size is set to 8, while the hyper-parameter of $\alpha$, $\beta$, and $\gamma$ are respectively 0.1, 2e-4, and 0.1 in default. When our framework uses MR-SSIM in the reconstruction loss and multiple description distance loss, the proposed method is marked as "Ours-mr". But, the proposed method is denoted as "Ours-ms", if MS-SSIM is used for reconstruction loss and multiple description distance loss.

\subsection{The objective and visual quality comparisons of different methods}

\begin{figure*}[!htb]
\centering
\includegraphics[width=6in]{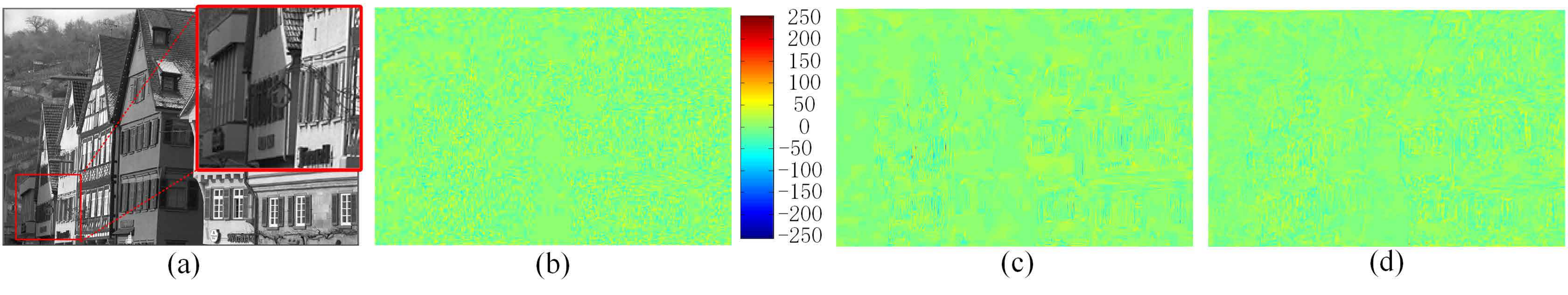}
\caption{The difference comparisons between two decoded side images of several multiple description coding methods. (a) one testing image from the Kodak PhotoCD dataset, (b-d) are respectively the difference images for "tcsvt18" (0.25 bpp(bits-per-pixel)) \cite{qq11}, "tip14" (0.29 bpp) \cite{qq13} and our method (0.24 bpp)}
\label{Fig3}
\end{figure*}
\begin{figure*}[!htb]
\centering
\includegraphics[width=6in]{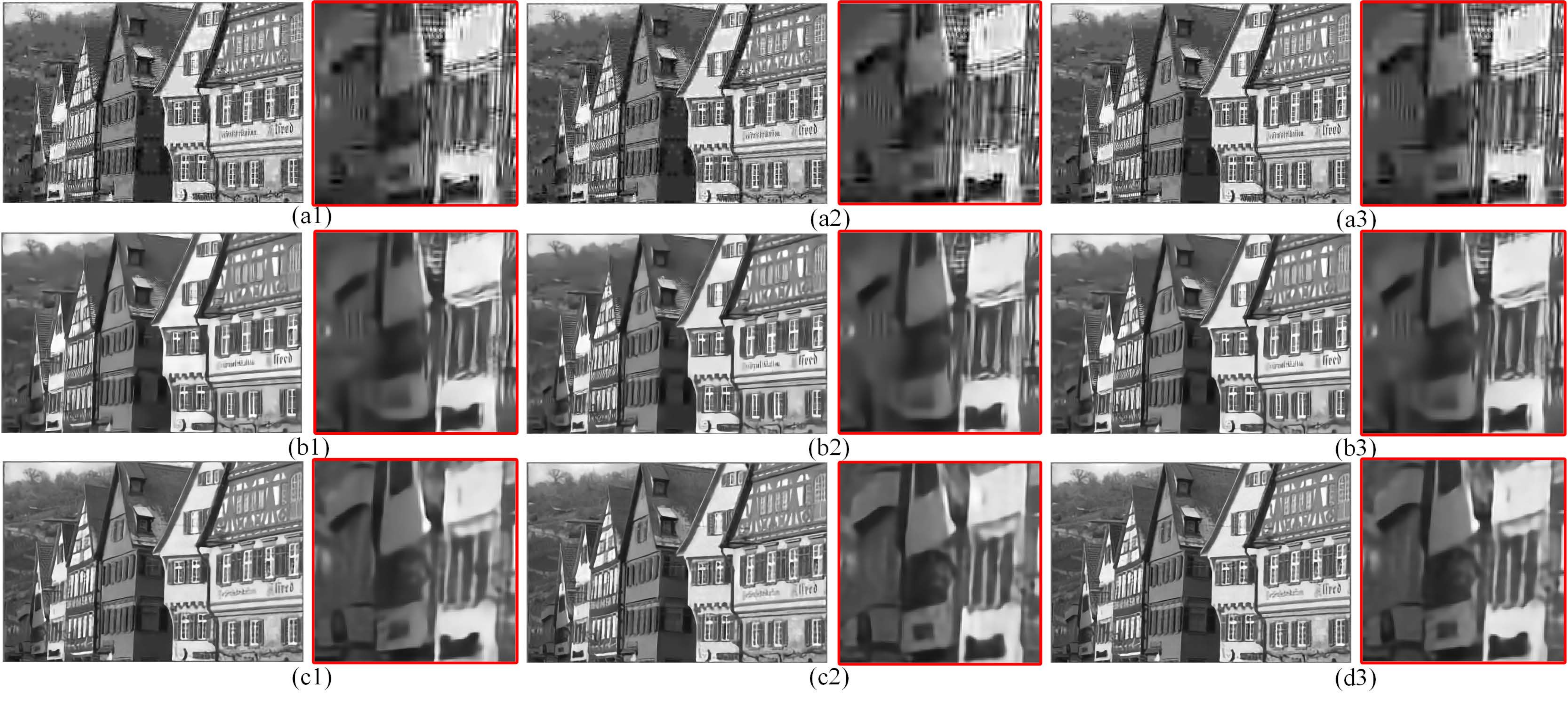}
\caption{The visual quality comparisons of different multiple description coding methods. (a1-a3),(b1-b3) and (c1-c3) are two decoded side images and central image respectively with "tcsvt18" (0.25 bpp) \cite{qq11}, "tip14" (0.29 bpp) \cite{qq13}, our method (0.24 bpp)}
\label{Fig4}
\end{figure*}
To validate the efficiency of the proposed framework, we compare our method with the latest standard-compatible CNN-based MDC method \cite{qq11} and multiple description coding approach with randomly offset quantizers \cite{qq13}, which are respectively denoted as "tcsvt18" \cite{qq11} and "tip14" \cite{qq13}. As described in \cite{qq21}, SSIM is a good approximation to assess image quality from the aspect of human visual perception, but this method only considers image single-scale information. Compared with SSIM, MS-SSIM is an image synthesis approach for image quality assessment, which considers distorted image's relative importance across different scales. Consequently, both MS-SSIM and MR-SSIM are chosen as the objective measurements to the assess distorted image's quality. Note that each scale SSIM weight factor of MR-SSIM is proportional to image size, but MS-SSIM's weights are obtained according to visual testing. At last, the visual comparisons of different MDC methods are given to observe the image quality, because human eyes are the ultimate recipient of the compressed images.

Although "Ours-mr" is trained with MR-SSIM instead of MS-SSIM, both of MR-SSIM and MS-SSIM results of several MDC approaches are shown in Fig. \ref{Fig2}, testing on two datasets. Meanwhile, we also show the MR-SSIM and MS-SSIM results of "Ours-ms". From this figure, we can see that "Ours-mr" has better performance than "Ours-ms" regarding MR-SSIM and MS-SSIM. Moreover, the objective MS-SSIM measurements of side and central decoded images between "tcsvt18" and "tip14" are very similar testing on Set4, when the bit-rate is higher than about 0.3 bpp. However, "tip14" has better performance than "tcsvt18" at very low bit-rate. The MR-SSIM measurements of side and central decoded images for "tip14" are always higher than "tcsvt18" testing on Set4. Although the coding efficiency of "tip14" is higher than "tcsvt18" in terms of MR-SSIM and MS-SSIM for side and central decoding when testing on the Kodak PhotoCD dataset at very low bit-rate, the side and central decoded images of "tip14" has less objective measurement values than the ones of "tcsvt18" regarding MR-SSIM and MS-SSIM when the costed bpp for multiple description coding is between 0.3 and 0.55. As compared to "tcsvt18" and "tip14", "Ours-mr" has the best coding efficiency on all the testing datasets in terms of MR-SSIM and MS-SSIM for both side and central decoded images, as depicted in Fig. \ref{Fig2}. We can also find that the coding efficiency of "Ours-mr" is far higher than two comparative methods at low bit-rate when testing on public image compression datasets.

As shown in Fig. \ref{Fig4} (a-c), "Ours-mr" can keep more structures of each object than "tcsvt18" \cite{qq11} and "tip14" \cite{qq13} for both of side and central decoded images. Although "tcsvt18" \cite{qq11} can preserve some small structures, this method makes many significant objects disappear, as displayed in Fig. \ref{Fig4} (b1-b3). Note that the shape of some objects is greatly distorted when testing images are compressed by "tcsvt18" \cite{qq11} and "tip14" \cite{qq13}, which can be seen in Fig. \ref{Fig4} (b1-b3, a1-a3). Moreover, "Ours-mr" does not have obvious visual noises such as coding artifacts, and the decoded side and central images look more natural, as compared to "tip14" \cite{qq13}. Besides, we also give the visual comparison of the differences between two decoded side images with different MDC approaches, as shown in Fig. \ref{Fig3}, from which we can see that the difference image for "tip14" \cite{qq13} includes many coding artifacts differences and the difference image of "tcsvt18" has many small structural differences \cite{qq11}, since the method of "tcsvt18" is trained with single-scale structural dissimilarity loss. However, "Ours-mr" has more structural differences on each scales rather than single scale, since our framework is trained with combinatorial structural dissimilarity loss on three-scales.

\section{Conclusion}
In this paper, we propose a deep multiple description coding framework, which includes multiple description multi-scale dilated encoder network, multiple description decoder networks, multiple description scalar quantization, and context-based entropy estimation neural network. Our multiple description quantizers are automatically learned by training a deep multiple description coding framework. Through testing on commonly used data-sets, it has been verified that our method has better coding efficiency than several advanced multiple description image compression methods, especially at low bit-rates.

\bibliographystyle{IEEEbib}
\bibliography{Template}

\begin{thebibliography}{10}

\bibitem{qq27}
V.~A. Vaishampayan,
\newblock ``{Design of multiple description scalar quantizers},''
\newblock {\em IEEE Transactions on Information Theory}, vol. 39, no. 3, pp.
  821--834, 1993.

\bibitem{qq17}
H.~Wu, T.~Zheng, and S.~Dumitrescu,
\newblock ``{On the design of symmetric entropy-constrained multiple
  description scalar quantizer with linear joint decoders},''
\newblock {\em IEEE Transactions on Communications}, vol. 65, no. 8, pp.
  3453--3466, 2017.

\bibitem{qq28}
Goyal~V. K.,
\newblock ``{Scalar quantization with random thresholds},''
\newblock {\em IEEE Signal Processing Letters}, vol. 18, no. 9, pp. 525--528,
  2011.

\bibitem{qq13}
L.~Meng, J.~Liang, U.~Samarawickrama, Y.~Zhao, H.~Bai, and A.~Kaup,
\newblock ``Multiple description coding with randomly and uniformly offset
  quantizers,''
\newblock {\em IEEE Transactions on Image Processing}, vol. 23, no. 2, pp.
  582--95, 2014.

\bibitem{qq10}
S.~Dumitrescu and Y.~Wan,
\newblock ``{Bit-error resilient index assignment for multiple description
  scalar quantizers},''
\newblock {\em IEEE Transactions on Information Theory}, vol. 61, no. 5, pp.
  2748--2763, 2015.

\bibitem{qq26}
H.~Jafarkhani and V.~Tarokh,
\newblock ``{Multiple description trellis coded quantization},''
\newblock in {\em IEEE International Conference on Image Processing}, Chicago,
  Oct. 1998.

\bibitem{qq6}
C.~Lin, Y.~Zhao, and C.~Zhe,
\newblock ``{Two-stage multiple description image coding using TCQ},''
\newblock {\em International Journal of Wavelets Multiresolution and
  Information Processing}, vol. 7, no. 5, pp. 665--673, 2009.

\bibitem{qq14}
V.~Vaishampayan, N.~Sloane, and S.~Servetto,
\newblock ``Multiple-description vector quantization with lattice codebooks:
  design and analysis,''
\newblock {\em IEEE Transactions on Information Theory}, vol. 47, no. 5, pp.
  1718--1734, 2001.

\bibitem{qq15}
V.~Goyal, J.~Kelner, and J.~Kovacevic,
\newblock ``Multiple description vector quantization with a coarse lattice,''
\newblock {\em IEEE Transactions on Information Theory}, vol. 48, no. 3, pp.
  781--788, 2002.

\bibitem{qq18}
S.~Dumitrescu, Y.~Chen, and J.~Chen,
\newblock ``{Index mapping for bit-error resilient multiple description lattice
  vector quantizer},''
\newblock {\em IEEE Transactions on Communications}, vol. PP, no. 99, pp. 1--1,
  2018.

\bibitem{qq7}
Z.~Gao and S.~Dumitrescu,
\newblock ``Flexible multiple description lattice vector quantizer with $l\geq
  3$ descriptions,''
\newblock {\em IEEE Transactions on Communications}, vol. 62, no. 12, pp.
  4281--4292, 2014.

\bibitem{qq11}
L.~Zhao, H.~Bai, A.~Wang, and Y.~Zhao,
\newblock ``{Multiple description convolutional neural networks for image
  compression},''
\newblock {\em IEEE Transactions on Circuits and Systems for Video Technology},
  vol. PP, no. 99, pp. 1--1, 2018.

\bibitem{qq22}
M.~Fabian, A.~Eirikur, T.~Michael, T.~Radu, and V.~G. Luc,
\newblock ``{Conditional probability models for deep image compression},''
\newblock in {\em IEEE Conference on Computer Vision and Pattern Recognition},
  Salt Lake City, June 2018.

\bibitem{qq41}
E.~Agustsson, F.~Mentzer, and M.~{Tschannen, et al.},
\newblock ``{Soft-to-hard vector quantization for end-to-end learned
  compression of images and neural networks},''
\newblock in {\em Advances in Neural Information Processing Systems},
  California, Dec. 2017.

\bibitem{qq20}
B.~Yochai and M.~Tomer,
\newblock ``{The perception-distortion tradeoff},''
\newblock in {\em IEEE Conference on Computer Vision and Pattern Recognition},
  Salt Lake City, June 2018.

\bibitem{qq21}
Z.~Wang, E.~P. Simoncelli, and A.~C Bovik,
\newblock ``{Multiscale structural similarity for image quality assessment},''
\newblock in {\em The Thrity-Seventh Asilomar Conference on Signals, Systems
  and Computers, 2003}, Pacific Grove, Nov. 2003.

\bibitem{qq23}
F.~Yu and V.~Koltun,
\newblock ``{Multi-scale context aggregation by dilated convolutions},''
\newblock in {\em arXiv:1511.07122}, 2015.

\bibitem{qq24}
P.~Wang, P.~Chen, D.~Liu Y.~Yuan, Z.Huang, X.~Hou, and G.Cottrell,
\newblock ``{Understanding convolution for semantic segmentation},''
\newblock in {\em IEEE Winter Conference on Applications of Computer Vision},
  Lake Tahoe, Mar. 2018.

\end{thebibliography}

\end{document}